\documentclass[aps,prl,superscriptaddress,showpacs,twocolumn]{revtex4}
\usepackage{amsmath,amssymb,revsymb,epsfig}

\usepackage{theorem}
\newtheorem{definition}{Definition}

\newtheorem{lemma}[definition]{Lemma}

\newlength{\blank}
\newenvironment{proof}[1][{\hspace{-\blank}}]{{\noindent\emph{Proof~{#1}.\ }}}{\hfill $\Box$\vskip 0.5\baselineskip}

\newcommand{\tr}{\operatorname{Tr}}
\newcommand{\ket}[1]{\left | #1 \right \rangle}
\newcommand{\bra}[1]{\left \langle #1 \right |}

\newcommand{\hilbert}{{\cal H}}

\begin{document}

\title{The entangling and disentangling power of unitary transformations are unequal}

\author{Noah Linden}
\email{n.linden@bristol.ac.uk}
\affiliation{Department of Mathematics, University of Bristol, Bristol BS8 1TW, United Kingdom}

\author{John A. Smolin}
\email{smolin@watson.ibm.com}
\affiliation{IBM T. J. Watson Research Center, Yorktown Heights, NY 10598, USA}

\author{Andreas Winter}
\email{a.j.winter@bris.ac.uk}
\affiliation{Department of Mathematics, University of Bristol, Bristol BS8 1TW, United Kingdom}

\date{22th November 2005}

\begin{abstract}
  We consider two capacity quantities associated with bipartite unitary
  gates: the entangling and the disentangling power.
  For two-qubit unitaries these two capacities are always the
  same.  Here we prove that these capacities are different
  in general.  We do so by constructing an explicit example of a qubit-qutrit unitary whose
  entangling power is maximal (2 ebits), but whose disentangling power
  is strictly less.  A corollary is that there can be no unique
  ordering for unitary gates in terms of their ability to perform
  non-local tasks.  Finally we show that in large dimensions, almost all
  bipartite unitaries have entangling and disentangling capacities close to the maximal
  possible (and thus in high dimensions the difference in these capacities is small for almost
  all unitaries).
\end{abstract}


\maketitle


{\bf Introduction.} In quantum information theory, we wish to
understand and compare states, channels or interactions via their
usefulness at certain operational tasks: creation of EPR states,
communication of classical or quantum bits, etc. Bipartite
unitaries have proved to be a fruitful arena in which to study
interactive communication tasks~\cite{Collins01,Eisert00,Chefles00,Zanardi00,2x2-form,CiracDur01,BerrySanders02,
BerrySanders03,BHLS03,LHL03,Nielsen03}.  In particular, the degree of
interaction of a bipartite unitary may be quantified in a number
of ways: for example its ability to perform forward or backward
communication, its ability to simulate other interactions, or, as
is of interest to us here, its ability to increase or decrease the
entanglement between two parties.

Formally,
we consider a unitary transformation $U$ acting on a bipartite
system shared by two observers Alice and Bob.   Alice has a system
Hilbert space $\mathcal{H}_A$ and Bob a system Hilbert space
$\mathcal{H}_B$.  The unitary $U=U_{AB}$ acts on
$\mathcal{H}_A\otimes\mathcal{H}_B$.  Alice (resp. Bob) also has
an ancilla with Hilbert space $\mathcal{H}_a$ (resp. $\mathcal{H}_b$).
Thus we extend the action of $U$ to the full Hilbert space as
$I_a\otimes U_{AB} \otimes I_b$ where $I_a$ (resp. $I_b$) is the
identity operator on $\mathcal{H}_a$ (resp. $\mathcal{H}_b$).  We
consider an initial state $\ket{\Psi^{\rm in}}$ on the full Hilbert
space, then act with $I_a\otimes U_{AB} \otimes I_b$ to produce a final
state
\begin{equation}
  \label{eq:in-out}
  \ket{\Psi^{\rm out}}=I_a\otimes U_{AB} \otimes I_b\ket{\Psi^{\rm in}}.
\end{equation}
Let $E(\Psi^{\rm in})$ be the entanglement of $\ket{\Psi^{\rm in}}$,
measured by the entropy of its reduced state
on the space $\mathcal{H}_A\otimes\mathcal{H}_a$~\cite{entanglement}.
Then the entangling power of $U$, which we denote $E^\uparrow(U)$ is
defined to be the maximum possible increase in the entanglement as
the input state varies:
\begin{equation}
  \label{eq:E-up}
  E^\uparrow(U)=\sup_{\ket{\Psi^{\rm in}}} \Bigl( E(\Psi^{\rm out}) - E(\Psi^{\rm in}) \Bigr).
\end{equation}

The disentangling power of $U$, denoted $E^\downarrow(U)$ is the
maximum decrease in entanglement that $U$ can effect:
\begin{equation}
  \label{eq:E-down}
  E^\downarrow(U)=\sup_{\ket{\Psi^{\rm in}}} \Big(E(\Psi^{\rm in}) - E(\Psi^{\rm out}) \Bigr).
\end{equation}

Clearly, $E^\downarrow(U)=E^\uparrow(U^\dagger)$. Note that by the
results of \cite{BHLS03,LHL03}, $E^\uparrow(U)$ is equal to the
asymptotic (many copies of $U$) capacity of $U$ to generate
entanglement (measured by the rate of EPR states) -- in these papers
it is shown that this capacity is given by the
single-letter formula eq.~(\ref{eq:E-up}), and that it is sufficient
to do the optimisation over pure states.

In this paper, we prove that in general $E^\uparrow(U)$ and
$E^\downarrow(U)$ are not equal. We show this by constructing and
analyzing an explicit example in $2\times 3$ dimensions. It is
worth recalling \cite{BerrySanders03}, for $2\times 2$--unitaries, that
$E^\uparrow(U)=E^\downarrow(U)=E^\uparrow(U^\dagger)$; thus our example
occurs in the smallest possible dimension.

It will be noticed that we have not said anything up to this point
about the relative dimensions of the system and ancilla Hilbert
spaces.  It is known that for typical unitaries $U$ it is
essential to have ancillas in order to generate the maximum
possible entanglement using $U$.  A well-known extreme case is the
SWAP operation on two qubits: it generates no entanglement
increase if Alice and Bob each only have the qubit on which the
SWAP acts, but it generates two ebits, the maximum possible
increase for any unitary acting on two qubits, if Alice and Bob
each have a qubit ancilla (i.e. Alice and Bob's local Hilbert
space are each two qubits).

For an arbitrary $U$ it is not known what size the
ancillas need to be to reach the maximum possible entanglement
increase (or decrease) for that unitary, or if indeed a maximiser
exists in finite dimension. Until now this has been a major
stumbling block in the calculation of the non-local capacities of
interactions~\cite{BHLS03,LHL03}.

{\bf Two-qubit gates~\cite{BerrySanders03}.}
It is well-known that a unitary of two qubits can, up to local unitary
equivalence, be written~\cite{2x2-form} as
\begin{equation}
  U = \exp\bigl( i\alpha\sigma_x\!\otimes\!\sigma_x
                +i\beta\sigma_y\!\otimes\!\sigma_y +i\gamma\sigma_z\!\otimes\!\sigma_z \bigr),
\end{equation}
with real numbers $\alpha$, $\beta$ and $\gamma$.
Consider any input state $\ket{\Psi^{\rm in}}$ and output state
\[
  \ket{\Psi^{\rm out}}
        = \exp\bigl( i\alpha\sigma_x\!\otimes\!\sigma_x
                    +i\beta\sigma_y\!\otimes\!\sigma_y
                    +i\gamma\sigma_z\!\otimes\!\sigma_z \bigr) \ket{\Psi^{\rm in}}.
\]
Then, taking the unitary to the other side and considering the
complex conjugate, we get
\[
  \ket{\overline{\Psi^{\rm in}}}
        = \exp\bigl( i\alpha\sigma_x\!\otimes\!\sigma_x
                    +i\beta\sigma_y\!\otimes\!\sigma_y
                    +i\gamma\sigma_z\!\otimes\!\sigma_z \bigr) \ket{\overline{\Psi^{\rm out}}},
\]
because the $\sigma_j\otimes\sigma_j$ are all real symmetric.
Since for all states, $E(\Psi) = E(\overline{\Psi})$, we can match
any entanglement increase (for input state $\ket{\Psi^{\rm in}}$)
by an equal entanglement decrease (for input state
$\ket{\overline{\Psi^{\rm out}}}$).

{\bf Main result.}
Our proof that the entangling and disentangling power are unequal
proceeds in two steps.  First we show that if a unitary
transformation has the largest possible entangling power for a
unitary of that dimension, then the local ancillas need only be
as large as the local system Hilbert spaces
(Lemma~\ref{lemma:max} below).
Then we exhibit an explicit unitary
acting on  ${\mathbb C}^2\otimes{\mathbb C}^3$ which has the
property that its entangling power is maximal (i.e. 2 ebits) but
its disentangling power is strictly less than 2 ebits, which we
prove by contradiction using our Lemma~\ref{lemma:max}.

\begin{lemma}
  \label{lemma:max}
  Let $U$ be a unitary acting on ${\mathbb C}^A\otimes{\mathbb C}^B$
  (without lost of generality we assume that $A\leq B$). If
  $U$ is maximally entangling (i.e., $E^\uparrow(U)=2\log A$~\cite{footnote-max}),
  then in eqs.~(\ref{eq:in-out}) and (\ref{eq:E-up}) one may restrict to
  ancillas of dimension $a=A$ and $b=B$; in particular, the supremum is a
  maximum, achieved using an input state of the product form
  \[
    \ket{\Psi^{\rm in}}_{aABb} = \ket{\Phi}_{aA}\otimes\ket{\Psi}_{Bb},
  \]
  with $\ket{\Phi}_{aA} = \frac{1}{\sqrt{A}}\sum_{j=1}^A \ket{j}_a\ket{j}_A$
  a maximally entangled state on $a\times A$
  and some $\ket{\Psi}_{Bb}$ on $B\times b$.
\end{lemma}
\begin{proof}
First, assume that for some ancillas of size $a$ and $b$,
respectively, there is actually a maximizer $\ket{\Psi^{\rm in}}$
-- after this we will give a proof that avoids this unwarranted
assumption. Generally, subadditivity of entropy~\cite{Wehrl}
implies the entanglement of the final state $E(\Psi^{\rm out})$
satisfies
\begin{eqnarray}
  E(\Psi^{\rm out}) = S(\rho_{Aa}^{\rm out})\leq S(\rho_{a}) + \log A.
  \label{ineq-a}
\end{eqnarray}
Also, the triangle inequality~\cite{Wehrl} implies that
\begin{eqnarray}
  E(\Psi^{\rm in}) = S(\rho_{Aa}^{\rm in})\geq S(\rho_{a}) - \log A.
  \label{ineq-b}
\end{eqnarray}
(Notice that since the unitary $U$ does not act on the ancilla
Hilbert space, $S(\rho_{a})$ is the same before and after the
action of $U$.) Thus,
\begin{eqnarray}
  \label{eq:gap}
  E(\Psi^{\rm out}) - E(\Psi^{\rm in})\leq 2\log A,
\end{eqnarray}
but since we assumed that
$E(\Psi^{\rm out}) - E(\Psi^{\rm in})= 2\log A$, we
must have equality in eqs.~(\ref{ineq-a}) and (\ref{ineq-b}).

Now we can calculate, using the above and the purity of the
state of four parties,
\begin{equation*}\begin{split}
  S(\rho_{ABb}^{\rm in}) &= S(\rho_a)                      \\
                         &= S(\rho_{Aa}^{\rm in}) + \log A \\
                         &= S(\rho_{Bb}^{\rm in}) + \log A.
\end{split}\end{equation*}
Thus we must have
\begin{eqnarray}
  \rho_{ABb}^{\rm in} = \frac{1}{A}I_A\otimes\rho_{Bb}^{\rm in},
  \label{equality-1}
\end{eqnarray}
and we may purify the  state $ \rho_{ABb}^{\rm in}$ by
writing Alice's ancilla Hilbert space in the form
$\hilbert_{a_1}\otimes\hilbert_{a_2}$,
so that the state of the full system is
\begin{eqnarray}
  \ket{\Psi^{\rm in}}_{aABb} = \ket{\Psi_1^{\rm in}}_{a_1A}\otimes\ket{\Psi_2^{\rm in}}_{a_2Bb}.
  \label{equality-2}
\end{eqnarray}
We may take $a_1$ to have dimension $A$ and $\ket{\Psi_1^{\rm
in}}_{Aa_1}$ is maximally entangled,
and hence $\rho_{a_1} = \frac{1}{A} I_{a_1}$.

We now consider the state after the action of $U$.
Eq.~(\ref{ineq-a}) with equality means that
$\rho_{Aa}^{\rm out} = \frac{1}{A}I_A \otimes \rho_{a}$,
so that
\begin{equation}\begin{split}
  \rho_{Aa_1a_2}^{\rm out} &= \frac{1}{A}I_A \otimes \rho_a \\
                           &= \frac{1}{A}I_A \otimes \frac{1}{A}I_{a_1}\otimes\rho_{a_2}.
  \label{equality-3}
\end{split}\end{equation}
Hence,
\begin{align*}
  E(\Psi^{\rm in}) &= S(\rho_{a_2}),          \\
  E(\Psi^{\rm out})&= S(\rho_{a_2})+ 2 \log A,
\end{align*}
from eqs.~(\ref{equality-2}) and (\ref{equality-3}).

We may now see that there is a different initial state which
yields the same entanglement increase.  We take exactly the
state (\ref{equality-2}) but now consider the situation in which the
ancilla particle $a_2$ is transferred to Bob --- let us relabel
$\tilde{a}=a_1$ and $\tilde{b}=ba_2$. Thus consider the initial state
\begin{eqnarray}
  \ket{\tilde\Psi^{\rm in}}_{\tilde{a}AB\tilde{b}}
             = \ket{\Psi_1^{\rm in}}_{\tilde{a}A} \otimes \ket{\Psi_2^{\rm in}}_{B\tilde{b}}.
\end{eqnarray}
This state has
\begin{align*}
  E(\tilde\Psi^{\rm in}) &= 0,       \\
  E(\tilde\Psi^{\rm out})&= 2 \log A,
\end{align*}
and we are done.

However, as we have said at the beginning of this proof, we cannot
assume that the supremum in eq.~(\ref{eq:E-up}) is a maximum.
Instead, for every $\epsilon>0$ there exist ancilla dimensions
$a$ and $b$ and an initial state such that
\begin{equation}
  \label{eq:E-gap-eps}
  2\log A - \epsilon \leq E(\Psi^{\rm out}) - E(\Psi^{\rm in}) \leq 2\log A.
\end{equation}
For eqs.~(\ref{ineq-a}) and (\ref{ineq-b}) this implies
\begin{align}
  \label{ineq-a-eps}
  S(\rho_a) + \log A - \epsilon &\leq E(\Psi^{\rm out}) \leq S(\rho_a) + \log A, \\
  \label{ineq-b-eps}
  S(\rho_a) - \log A            &\leq E(\Psi^{\rm in})  \leq S(\rho_a) - \log A + \epsilon.
\end{align}
Eq.~(\ref{ineq-b-eps}) gives that for the mutual information of the
initial state between $A$ and $Bb$,
\begin{equation*}\begin{split}
  I(A:Bb)_{\Psi^{\rm in}}
                &=    S(\rho^{\rm in}_A) + S(\rho^{\rm in}_{Bb}) - S(\rho^{\rm in}_{ABb}) \\
                &=    S(\rho^{\rm in}_A) + E(\Psi^{\rm in}) - S(\rho_a)                   \\
                &\leq \log A             + S(\rho_a)-\log A+\epsilon - S(\rho_a)
                 =    \epsilon.
\end{split}\end{equation*}
Since the mutual information can be expressed by means of the
relative entropy,
\[
  I(A:Bb)_{\Psi^{\rm in}}
     = D\bigl( \rho^{\rm in}_{ABb} \| \rho^{\rm in}_A\otimes\rho^{\rm in}_{Bb} \bigr),
\]
and with Pinkser's inequality,
\[
  D(\rho\|\sigma) \geq \left( \frac{1}{2} \| \rho-\sigma \|_1 \right)^2,
\]
we find that
\begin{equation}
  \label{eq:close-to-product}
  \left\| \rho^{\rm in}_{ABb} - \rho^{\rm in}_A\otimes\rho^{\rm in}_{Bb} \right\|_1
                                                                \leq 2\sqrt{\epsilon}.
\end{equation}
The second state is a product state, so it has a purification of
product form, $\ket{\widehat\Psi_1^{\rm
in}}_{a_1A}\otimes\ket{\widehat\Psi_2^{\rm in}}_{a_2Bb}$, and by
Uhlmann's theorem~\cite{uhlmann} one can isometrically map $a$ to
$a_1 a_2$ such that
\begin{equation}
  \label{eq:pure:close-to-product}
  \left\| \ket{\Psi^{\rm in}}_{a_1a_2ABb} \!-\!
            \ket{\widehat\Psi_1^{\rm in}}_{a_1A} \!\otimes\!
                         \ket{\widehat\Psi_2^{\rm in}}_{a_2Bb} \right\|_1
                                                       \!\leq 2\sqrt[4]{\epsilon} =: \delta,
\end{equation}
where we have used a well-known relation between fidelity and trace
distance~\cite{Fuchs:vandeGraaf}.
%
Now we will switch over to
\[
  \ket{\widehat\Psi^{\rm in}}_{a_1a_2ABb}
       = \ket{\widehat\Psi_1^{\rm in}}_{a_1A}\otimes\ket{\widehat\Psi_2^{\rm in}}_{a_2Bb}
\]
as the new input state. Notice that, because $a$ is not affected by the dynamics,
\begin{equation}\begin{split}
  \label{eq:Psi}
  E(\Psi^{\rm out}) - E(\Psi^{\rm in}) &= S(Aa)_{\Psi^{\rm out}}-S(Aa)_{\Psi^{\rm in}}  \\
                                       &= S(A|a)_{\Psi^{\rm out}}-S(A|a)_{\Psi^{\rm in}},
\end{split}\end{equation}
and likewise for the new state $\widehat\Psi$:
\begin{equation}\begin{split}
  \label{eq:Psi-hat}
  E\bigl(\widehat\Psi^{\rm out}\bigr) - E\bigl(\widehat\Psi^{\rm in}\bigr)
                         &= S(Aa)_{\widehat\Psi^{\rm out}}-S(Aa)_{\widehat\Psi^{\rm in}}  \\
                         &= S(A|a)_{\widehat\Psi^{\rm out}}-S(A|a)_{\widehat\Psi^{\rm in}},
\end{split}\end{equation}
with the conditional entropy~\cite{Cerf:Adami}
$S(X|Y)_\rho = S(\rho_{XY})-S(\rho_Y)$.
This means we have only
to control how much the conditional entropy changes
when we modify the state, and this we can indeed do with the help of a
variant of Fannes' inequality, proved recently~\cite{Alicki:Fannes}:
for $\delta\leq 1/2$,
\[
  \Bigl| S(A|a)_{\widehat\Psi^{\rm in}} - S(A|a)_{\Psi^{\rm in}} \Bigr|
                                               \leq 2H_2(\delta)+4\delta\log A,
\]
and likewise for the output states, where we have used the binary
entropy $H_2(\delta)=-\delta\log\delta-(1-\delta)\log(1-\delta)$.
(Note that, unlike the usual Fannes inequality, we have only
a dependence on the dimension of $A$ but not of the ancilla.)

But now we can perform the same trick as above: we look at the
new input state obtained from $\ket{\widehat\Psi^{\rm in}}_{a_1a_2ABb}$
by handing $a_2$ to Bob, i.e., with $\tilde{a}=a_1$ and
$\tilde{b}=ba_2$,
\[
  \ket{\tilde\Psi^{\rm in}}_{\tilde{a}AB\tilde{b}}
       = \ket{\widehat\Psi_1^{\rm in}}_{a_1A}\otimes\ket{\widehat\Psi_2^{\rm in}}_{Bba_2}.
\]
Since both input and output state, restricted to $a_1a_2A$, are then
products across $a_1A$-$a_2$, we get
\begin{equation*}\begin{split}
  E\bigl(\tilde\Psi^{\rm out}\bigr) - E\bigl(\tilde\Psi^{\rm in}\bigr)
   &= S\bigl(\tilde\rho^{\rm out}_{Aa_1}\bigr)-S\bigl(\tilde\rho^{\rm in}_{Aa_1}\bigr)      \\
   &= S\bigl(\tilde\rho^{\rm out}_{Aa_1a_2}\bigr)-S\bigl(\tilde\rho^{\rm in}_{Aa_1a_2}\bigr)\\
   &= E\bigl(\widehat\Psi^{\rm out}\bigr) - E\bigl(\widehat\Psi^{\rm in}\bigr)              \\
   &\geq 2\log A - \epsilon - 4H_2(\delta) - 8\delta\log A.
\end{split}\end{equation*}
At this point we can perform the limit $\epsilon$ (and hence $\delta$)
$\rightarrow 0$: since $\tilde{a}$ purifies $A$ and $\tilde{b}$
purifies $B$, we can restrict their dimensions to $A$ and $B$,
respectively, and so the states $\ket{\tilde\Psi^{\rm in}}$ have
an accumulation point for which the difference between the output
and the input entanglement is precisely $2\log A$. This state
is a product state between Alice and Bob, and
just as in the earlier argument, it is now immediate
that Alice must have a maximally entangled state between
$A$ and her ancilla.
\end{proof}

We thus conclude that if a unitary creates the maximal amount of
entanglement, it can do so by acting on a state which is product
pure state between Alice and Bob. Furthermore the state on Alice's
side may be taken to be maximally entangled between the system and
ancilla. If $\hilbert_A$ and $\hilbert_B$  have different
dimensions (with $\hilbert_A$ assumed to be smaller) we may only
conclude that Bob's initial state may be taken to be pure with an
ancilla of the same size as $\hilbert_B$.

Of course if the
dimensions of $\hilbert_A$ and $\hilbert_B$ are equal then we may
run the argument again, with the roles of Alice and Bob
interchanged, to show that the initial state may be taken
to be a product state between Alice and Bob, with both Alice and
Bob maximally entangled with their local ancilla.
From this it is not hard to show that in this case,
still assuming that $U$ is maximally
entangling, it is also maximally disentangling. In other words,
for $A=B$,
\[
  E^\uparrow(U) = 2\log A \ \Longleftrightarrow\  E^\downarrow(U) = 2\log A.
\]

We now exhibit an explicit unitary transformation acting on
${\mathbb C}^2\otimes{\mathbb C}^3$ which can
entangle maximally, but which cannot disentangle maximally:
consider
\begin{equation}\begin{split}
  U_{2\times3} &= -i\ket{w_{00}}\bra{00} + \ket{w_{01}}\bra{01} + \ket{w_{02}}\bra{02} \\
               &\phantom{=-}
                 + \ket{w_{10}}\bra{10} + \ket{w_{11}}\bra{11} -i \ket{w_{12}}\bra{12},
\end{split}\end{equation}
with (for $j=0,1,2$)
\begin{align*}
  \ket{w_{0j}} &= \frac{1}{\sqrt 3}
                  \bigl( \ket\alpha\ket 0 + \omega^{j}\ket\beta\ket 1
                                           + \omega^{2j}\ket\gamma\ket 2 \bigr),             \\
  \ket{w_{1j}} &= \frac{1}{\sqrt 3}
                  \bigl( \ket{\alpha^\bot}\ket 0 + \omega^{j}\ket{\beta^\bot}\ket 1
                                                  + \omega^{2j}\ket{\gamma^\bot}\ket 2 \bigr).
\end{align*}
Here $\omega=e^{2\pi i/3}$ is the cube-root of unity,
and $\ket\alpha$, $\ket\beta$, $\ket\gamma$ and
$\ket{\alpha^\bot}$, $\ket{\beta^\bot}$, $\ket{\gamma^\bot}$
are sets of ``trine'' states:
\begin{align}
  \ket\alpha &\!=\! \ket{0},
       & \ket{\alpha^\bot}&\!=\! \ket{1},                           \nonumber\\
  \ket\beta  &\!=\!-\frac{1}{2}\ket{0}\!+\!\frac{\sqrt 3}{2}\ket{1},
       & \ket{\beta^\bot} &\!=\!-\frac{1}{2}\ket{1}\!-\!\frac{\sqrt 3}{2}\ket{0},
                                                                    \nonumber\\
  \ket\gamma &\!=\!-\frac{1}{2}\ket{0}\!-\!\frac{\sqrt 3}{2}\ket{1},
       & \ket{\gamma^\bot}&\!=\!-\frac{1}{2}\ket{1}\!+\!\frac{\sqrt 3}{2}\ket{0}.
                                                                    \nonumber
\end{align}

The unitary $U_{2\times3}$ can create two ebits.
For consider its action on the state
\begin{equation*}
  \ket{\Phi^{\rm in}_1} = \frac{1}{2}\bigl( {\ket 0}_a{\ket 0}_A+{\ket 1}_a{\ket 1}_A \bigr)
                              \otimes\bigl( {\ket 0}_B{\ket 0}_b+{\ket 2}_B{\ket 2}_b \bigr).
\end{equation*}
The subscript $A$ denotes Alice's system state and $a$ her ancilla
state; similarly $B$ denotes Bob's system state and $b$ his
ancilla state.  The unitary $U_{2\times3}$ acts on the Hilbert
spaces $A$ and $B$ (i.e. the full unitary is $I_a\otimes
U_{2\times 3}\otimes I_b$, where $I_a$ is the identity operator on
the $a$ Hilbert space). Clearly the initial state
$\ket{\Phi_1^{\rm in}}$ has zero entanglement between Alice and
Bob (i.e. between $Aa$ and $Bb$). It is not difficult to check
that the final state
\begin{eqnarray}
  \ket{\Phi^{\rm out}_1} = U_{2\times 3} \ket{\Phi^{\rm in}_1}
\end{eqnarray}
has entanglement of two ebits between Alice and Bob.
Thus $U_{2\times 3}$ has the maximum possible entangling power
for any unitary on ${\mathbb C}^2\otimes{\mathbb C}^3$.

We now show that the disentangling power of $U_{2\times 3}$
is strictly less than 2 ebits. It will be convenient
to analyze the entangling power of the inverse of $U_{2\times 3}$.
Assuming the contrary, by Lemma~\ref{lemma:max},
if $U_{2\times 3}$ would entangle maximally,
the state from which it creates most entanglement may be
taken to be a product state: $\ket{\eta_1}_{aA}\otimes \ket {\eta_2}_{Bb}$,
and without loss of generality $\ket{\eta_1}_{aA}$ is
a maximally entangled state of two qubits,
$\ket {\eta_2}_{Bb}$ an arbitrary pure state of
two qutrits.
Thus the most general input state we need to consider is
\begin{equation}\begin{split}
  \ket{\Phi^{\rm in}_2}
     &= \frac{1}{\sqrt 2}\bigl({\ket 0}_a{\ket 0}_A+{\ket 1}_a{\ket 1}_A\bigr)  \\
     &\phantom{==}
        \otimes\bigl( {\ket 0}_B{\ket {\tau_0}}_b
                     +{\ket 1}_B{\ket {\tau_1}}_b
                     +{\ket 2}_B{\ket {\tau_2}}_b \bigr).           \label{Psi-in2}
\end{split}\end{equation}
Normalization of $\ket{\Phi^{\rm in}_2}$ means that
\begin{eqnarray}
  \langle{\tau_0}\ket{\tau_0}_b
    +\langle{\tau_1}\ket{\tau_1}_b +\langle{\tau_2}\ket{\tau_2}_b = 1.
  \label{tau-normalization}
\end{eqnarray}
Clearly $\ket{\Phi^{\rm in}_2}$ has no entanglement between $Aa$
and $Bb$.

To compute the output state,
we begin by rewriting the inverse $U_{2\times 3}^\dagger$ as
\begin{equation*}\begin{split}
  U_{2\times 3}^\dagger
     &= {\ket 0}_A{\ket {v_0}}_B {\bra 0}_A {\bra 0}_B                                  \\
     &\phantom{=}
      + \left( -\frac{1}{2}{\ket 0}_A{\ket{v_1}}_B
               -\frac{\sqrt 3}{2} {\ket 1}_A{\ket{v_1'}}_B \right) {\bra 0}_A{\bra 1}_B \\
     &\phantom{=}
      + \left( -\frac{1}{2}{\ket 0}_A{\ket{v_2}}_B
               +\frac{\sqrt 3}{2} {\ket 1}_A{\ket{v_2'}}_B \right) {\bra 0}_A{\bra 2}_B \\
     &\phantom{.}
      + {\ket 1}_A{\ket {v_0'}}_B {\bra 1}_A {\bra 0}_B                                 \\
     &\phantom{=}
      + \left( \frac{\sqrt 3}{2}{\ket 0}_A{\ket{v_1}}_B
              -\frac{1}{2} {\ket 1}_A{\ket{v_1'}}_B \right) {\bra 1}_A{\bra 1}_B        \\
     &\phantom{=}
      + \left( -\frac{\sqrt 3}{2}{\ket 0}_A{\ket{v_2}}_B
               -\frac{1}{2} {\ket 1}_A{\ket{v_2'}}_B \right) {\bra 1}_A{\bra 2}_B,
\end{split}\end{equation*}
where (for $j=0,1,2$)
\begin{align*}
  \ket{v_j} &= \frac{1}{\sqrt 3}
                 \bigl( i\ket 0 + \omega^{-j} \ket 1 +  \omega^{-2j}\ket 2 \bigr), \\
  \ket{v_j'}&= \frac{1}{\sqrt 3}
                 \bigl(  \ket 0 + \omega^{-j} \ket 1 + i\omega^{-2j}\ket 2 \bigr).
\end{align*}
%
Thus the result of $U_{2\times 3}^\dagger$ acting on
(\ref{Psi-in2}) is
\begin{equation*}\begin{split}
  \ket{\Phi^{\rm out}_2} &= U_{2\times 3}^\dagger\ket{\Phi^{\rm in}_2}         \\
            &= \frac{1}{2}\Bigl[ {\ket 0}_a {\ket 0}_A {\ket {\Phi_{00}}}_{Bb}
                                 +{\ket 0}_a {\ket 1}_A {\ket {\Phi_{01}}}_{Bb} \Bigr. \\
            &\phantom{===}
                          \Bigl. +{\ket 1}_a {\ket 0}_A {\ket {\Phi_{10}}}_{Bb}
                                 +{\ket 1}_a {\ket 1}_A {\ket {\Phi_{11}}}_{Bb} \Bigr],
\end{split}\end{equation*}
where now
\begin{align*}
  \ket{\Phi_{00}} &= \sqrt 2\Bigl[ {\ket{v_0}}_{B}{\ket{\tau_{0}}}_{b}
                                       -\frac{1}{2} {\ket{v_1}}_{B}{\ket{\tau_{1}}}_{b}
                                       -\frac{1}{2} {\ket{v_2}}_{B}{\ket{\tau_{2}}}_{b}
                                 \Bigr],            \\
  \ket{\Phi_{01}} &= \sqrt 2\Bigl[ -\frac{\sqrt 3}{2}\ket{v_1'}_B\ket{\tau_{1}}_{b}
                                        +\frac{\sqrt 3}{2}\ket{v_2'}_B\ket{\tau_{2}}_{b}
                                 \Bigr],            \\
  \ket{\Phi_{10}} &= \sqrt 2\Bigl[ \frac{\sqrt 3}{2}\ket{v_1}_B\ket {\tau_{1}}_b
                                       -\frac{\sqrt 3}{2}\ket{v_2}_{B}\ket{\tau_{2}}_b
                                 \Bigr],            \\
  \ket{\Phi_{11}} &= \sqrt 2\Bigl[ \ket{v_0'}_B\ket{{\tau_{0}}}_b
                                       -\frac{1}{2}\ket{v_1'}_B\ket{\tau_{1}}_b
                                       -\frac{1}{2} {\ket {v_2'}}_{B}{\ket{\tau_{2}}}_{b}
                                 \Bigr].
\end{align*}
Now, in order for $\ket{\Phi^{\rm out}_2}$ to be maximally
entangled we require that the four states ${\ket {\Phi_{00}}}$,
${\ket {\Phi_{01}}}$, ${\ket {\Phi_{10}}}$ and ${\ket
{\Phi_{11}}}$ form an orthonormal basis:
\[
  \langle{\Phi_{ij}}{\ket {\Phi_{km}}}=\delta_{ij,km}.
\]
This puts constraints on the $\ket{\tau_j}$, which, as we shall
see, lead to a contradiction.

Bearing in mind the normalization of the $\ket{\tau_j}$,
eq.~(\ref{tau-normalization}), the four equations expressing the
condition that the vectors ${\ket {\Phi_{ij}}}$ be normalised are
all the same, namely:
\begin{equation*}
  \langle{\tau_1}\ket {\tau_1} +\langle{\tau_2}\ket {\tau_2} = \frac{2}{3},\
  \text{or equivalently, }\langle{\tau_0}\ket {\tau_0}       = \frac{1}{3}.
\end{equation*}
The requirement that $\langle \Phi_{00}{\ket {\Phi_{10}}}=0$ thus
leads to
\begin{eqnarray}
  \langle{\tau_0}\ket{\tau_0}
    = \langle{\tau_1}\ket{\tau_1}
    = \langle{\tau_2}\ket{\tau_2}
    = \frac{1}{3}.
  \label{condition-00-10}
\end{eqnarray}
The requirement that $\langle \Phi_{01}{\ket {\Phi_{10}}}=0$
yields
\begin{align*}
  -\langle{\tau_1}\ket{\tau_1} -\langle{\tau_2}\ket{\tau_2}
       &+ (1-\sqrt 3)\omega^2 \langle{\tau_1}\ket{\tau_2}  \\
       &+ (1+\sqrt 3)\omega   \langle{\tau_2}\ket{\tau_1} = 0.
\end{align*}
This has the unique solution
$\langle{\tau_1}\ket{\tau_2} = \frac{\omega}{3}$,
and with Cauchy-Schwarz and eq.~(\ref{condition-00-10}), this means that
\begin{eqnarray}
  \label{eq:1-2-proportional}
  \ket{\tau_2}={\omega}\ket{\tau_1}.
\end{eqnarray}
The requirement that $\langle \Phi_{00}{\ket {\Phi_{01}}}=0$ gives
\begin{eqnarray}
  -(1+\sqrt 3)\omega^2\langle{\tau_0}\ket{\tau_1}
  +(1-\sqrt 3)\omega\langle{\tau_0}\ket{\tau_2}             & & \nonumber\\
  +\frac{1}{2}\langle{\tau_1}\ket{\tau_1}
  -\frac{1}{2}(1+\sqrt 3)\omega^2\langle{\tau_1}\ket{\tau_2}& & \nonumber\\
  +\frac{1}{2}(1-\sqrt 3)\omega\langle{\tau_2}\ket {\tau_1}
  -\frac{1}{2}\langle{\tau_2}\ket {\tau_2}                 &=& 0 \nonumber
\end{eqnarray}
Using eqs.~(\ref{condition-00-10}) and (\ref{eq:1-2-proportional}),
this implies that
\begin{eqnarray}
  \label{eq:01:02}
  \langle{\tau_0}\ket {\tau_1} = -\frac{\omega}{6}\quad\text{and}\quad
  \langle{\tau_0}\ket {\tau_2} = -\frac{\omega^2}{6}.
\end{eqnarray}
But now, inserting eqs.~(\ref{condition-00-10}),
(\ref{eq:1-2-proportional}) and (\ref{eq:01:02}), we get
\begin{equation*}\begin{split}
  \langle \Phi_{00}{\ket {\Phi_{11}}}
      &= \frac{2}{3}\langle{\tau_0}\ket{\tau_0}
            -\frac{1}{3}(1\!+\!\sqrt 3)\omega^2\langle{\tau_0}\ket{\tau_1} \\
      &\phantom{=}
        -\frac{1}{3}(1\!-\!\sqrt 3)\omega\langle{\tau_0}\ket{\tau_2}
            -\frac{1}{3}(1\!-\!\sqrt 3)\omega\langle{\tau_1}\ket{\tau_0}   \\
      &\phantom{=}
        +\frac{1}{6}\langle{\tau_1}\ket{\tau_1 }
            +\frac{1}{6}(1\!+\!\sqrt 3)\omega^2\langle{\tau_1}\ket{\tau_2}
            +\frac{1}{6}\langle{\tau_2}\ket{\tau_2}                    \\
      &\phantom{=}
        -\frac{1}{3}(1\!+\!\sqrt 3)\omega^2\langle{\tau_2}\ket{\tau_0}
            + \frac{1}{6}(1\!-\!\sqrt 3)\omega\langle{\tau_2}\ket{\tau_1}  \\
      &= \frac{2}{3} \neq 0.
\end{split}\end{equation*}
Thus there is no choice of $\ket{\tau_0}$, $\ket{\tau_1}$,
$\ket{\tau_2}$ for ${\ket{\Phi_{ij}}}$ to form an orthonormal
basis. This is the desired contradiction, and we conclude that
$E^\downarrow(U) < 2 = E^\uparrow(U)$.

{\bf Further thoughts and conclusion.} We have found an example of
bipartite unitary of smallest possible dimension such that its
entangling and its disentangling power are different. This is a
striking result as it shows that there can be no unique ordering
of unitary gates with respect to their various capacities. For consider
$U_1=U_{2\times 3}$ and $U_2=U^\dagger_{2\times 3}$: $U_1$ has greater entangling capacity
than $U_2$; but $U_1$ has smaller disentangling capacity than $U_2$.
Note however, that our proof is only by contradiction, and that in
particular we show only that there is a difference between
$E^\uparrow$ and $E^\downarrow$ but not how large it is.

We have done some numerical work, which, for our gate $U_{2\times 3}$,
indicates that
\[
  2 - E^\downarrow(U_{2\times 3}) \approx 0.06.
\]
Furthermore, we tried to find the maximum
difference $E^\uparrow(U)-E^\downarrow(U)$ over all
$2\times 3$--gates $U$, which seems to be $\approx 0.13$, and in
general for randomly selected unitary,
the entangling and the disentangling power doesn't seem to
be much different. [See Fig 1.]

\begin{figure}
  \includegraphics[width=3in]{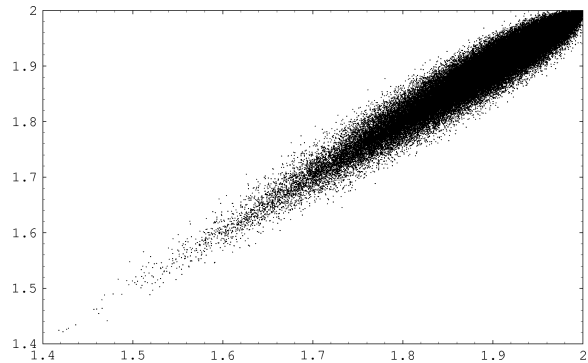}
  \caption{Entangling vs.~disentangling power for 170,000 randomly chosen $2\times 3$ unitaries.
  The local ancilla dimensions are 2 and 3 respectively.}
\end{figure}

We can explain this partly by the \emph{concentration
of measure phenomenon} in large dimensions [which usually
however kicks in for relatively small dimensions]: For a (random, according to
Haar measure) unitary $U_{AB}$ in $A\times B$ dimensions,
consider as initial state
\[
  \ket{\Phi^{\rm in}} = \ket{\Phi_A}_{aA}\otimes\ket{\Phi_B}_{Bb},
\]
the tensor product of two maximally entangled states between the
local systems and the respective local ancillas.  As before, we
will assume without loss of generality that $A\leq B$.
On the face of it, it might be thought that this
state may not be the best input state for a particular unitary, enabling us to achieve
the entangling capacity; and indeed it will not be for every unitary.  However, in fact, it
is a pretty good input state for most unitaries.  For we shall show
that the expected entanglement of $\ket{\Phi^{\rm out}}$ for this state is close
to $2\log A$; to be precise,
\begin{equation}
  \label{eq:expected-E}
  {\mathbb E}_U E(\Phi^{\rm out}) \geq 2\log A - \frac{1}{\ln 2}\frac{A^2}{B^2}.
\end{equation}
For this, we use the inequality
$S(\rho) = -\tr\rho\log\rho \geq -\log\tr\rho^2 = S_2(\rho)$
between the von Neumann and the R\'{e}nyi entropy, and then have
\begin{equation*}\begin{split}
  {\mathbb E}_U E(\Phi^{\rm out})
                   &\geq {\mathbb E}_U \Bigl[ -\log\tr\bigl(\rho^{\rm out}_{aA}\bigr)^2 \Bigr] \\
                   &\geq -\log {\mathbb E}_U \tr\bigl(\rho^{\rm out}_{aA}\bigr)^2,
\end{split}\end{equation*}
by the convexity of $-\log x$. To evaluate the quadratic average in the last
line, we rewrite it as follows:
\begin{equation*}\begin{split}
  \tr\bigl(\rho^{\rm out}_{aA}\bigr)^2
                   &= \tr\Bigl( \bigl(\rho^{\rm out}_{aA}\otimes\rho^{\rm out}_{aA}\bigr)
                                                                              F_{aA,aA} \Bigr) \\
                   &= \tr\Bigl( \bigl(\Phi^{\rm out}_{aABb}\otimes\Phi^{\rm out}_{aABb}\bigr)
                                                            (F_{aA,aA}\otimes I_{BbBb}) \Bigr),
\end{split}\end{equation*}
with the SWAP operator $F$ on $aA\,aA$. Hence
\begin{equation*}
  {\mathbb E}_U \tr\bigl(\rho^{\rm out}_{aA}\bigr)^2
                    = \tr\Bigl( \omega (F_{aA,aA}\otimes I_{BbBb}) \Bigr),
\end{equation*}
where --- with the symmetric and antisymmetric states
\begin{align*}
  \sigma_{AB} &= \frac{I_{ABAB}+F_{AB,AB}}{AB(AB+1)}
               = \frac{2}{AB(AB+1)}\Pi_{\rm sym} \text{ and} \\
  \alpha_{AB} &= \frac{I_{ABAB}-F_{AB,AB}}{AB(AB-1)}
               = \frac{2}{AB(AB-1)}\Pi_{\rm anti},
\end{align*}
respectively ---
\begin{equation*}\begin{split}
  \omega &= \bigl( {\rm id}_{aabb}\otimes {\cal T}_{AB,AB} \bigr)
                                            \bigl( \Phi^{\rm in}\otimes\Phi^{\rm in} \bigr) \\
         &= \frac{1}{2}\left( 1\!+\!\frac{1}{AB} \right)\sigma_{ab}\otimes\sigma_{AB}
            +\frac{1}{2}\left( 1\!-\!\frac{1}{AB} \right)\alpha_{ab}\otimes\alpha_{AB}.
\end{split}\end{equation*}
Here,
\begin{equation*}\begin{split}
  {\cal T}_{AB,AB}(\rho) &= \int {\rm d}U (U\otimes U)\rho(U\otimes U)^\dagger             \\
                         &= \sigma_{AB}\tr\bigl( \rho \Pi_{\rm sym} \bigr)
                           +\alpha_{AB}\tr\bigl( \rho \Pi_{\rm anti} \bigr)
\end{split}\end{equation*}
is the twirling operation~\cite{BDSW} --- the second line follows from the
fact that the $U\otimes U$-representation decomposes into the
symmetric and the antisymmetric part.
Using the formulas $\tr_{BB} I_{ABAB} = B^2 I_{AA}$ and
$\tr_{BB} F_{AB,AB} = B F_{A,A}$,
this yields straightforwardly
\[
  {\mathbb E}_U \tr\bigl(\rho^{\rm out}_{aA}\bigr)^2
                    = \frac{A^2+B^2-2}{A^2B^2-1} \leq \frac{1}{A^2}+\frac{1}{B^2},
\]
which immediately implies eq.~(\ref{eq:expected-E}). A deep result from
probability theory, \emph{Levy's lemma}
(see~\cite{Ledoux}), as applied in~\cite{aspects}, now informs us
that the probability of a random $U$ yielding less than $2\log A -
\frac{1}{\ln 2}\frac{A^2}{B^2} - \epsilon$ ebits for $\ket{\Phi^{\rm
out}}$ is bounded above by
$$\exp\left( -\frac{\text{const.}}{(\log A)^2} \epsilon^2 AB \right).$$
Since with $U$ also $U^\dagger$ is Haar distributed,
we conclude that (for large $A\leq B$ or
for $B$ much larger than $A$) a random $U$ is overwhelmingly likely to have
entangling and disentangling power close to the maximum $2\log A$
(and thus the difference between these capacities is also likely to be small).

We note however, that this does not preclude the possibility that a particular unitary
could have entangling and disentangling power very different from each other. [Although
our numerical evidence, described above, shows that this does not seem to happen in
dimension $2\times 3$ for example].
Indeed, independent work by Harrow and Shor~\cite{Aram:Peter}
shows that for large local dimensions $d$, it is possible to
construct a unitary for which $E^\uparrow(U) - E^\downarrow(U) \sim \log d$.

\medskip
\begin{acknowledgments}
  NL and AW  thank the EU for support through the European
  Commission project RESQ (contract IST-2001-37559);  NL, AW and JAS
  thank the UK EPSRC for support through the Interdisciplinary
  Research Collaboration in Quantum Information Processing; JAS
  thanks the US National Security Agency and the Advanced Research
  and Development Activity for support through contract
  DAAD19-01-C-0056. We also would like to thank A. Harrow and
  P. Shor for conversations about their work~\cite{Aram:Peter}.
\end{acknowledgments}


\begin{thebibliography}{10}
\bibitem{footnote-max} An easy way of seeing that $2\log A$ is the maximum entanglement
    is to note that the gate can be simulated using this amount of entanglement: Alice
    teleports her system to Bob, who applies the gate and teleports Alice's
    system back.
\end{thebibliography}
\end{document}